\documentclass{article}

\title{ \textbf{Bipartite Link Prediction \\based on Toplogical Features\\ via 2-hop Path}}
\author{\textbf{Jungwoon Shin} \\ Korea University \\ jungwoonshin@gmail.com} 
\date{February 2020}
\usepackage[utf8]{inputenc}
\usepackage{hyperref}
\usepackage{algorithm}
\usepackage{algorithmic}
\usepackage[margin=1in]{geometry}
\usepackage[table,xcdraw]{xcolor}
\usepackage{graphicx}
\usepackage{lscape}
\usepackage{adjustbox}
\usepackage{caption}
\usepackage{booktabs}
\usepackage{array}
\usepackage{float}
\usepackage{multirow}
\usepackage{color}
\usepackage{graphicx}
\graphicspath{ {.} }

\begin{document}
\maketitle
\nocite{*}
\begin{abstract}
A variety of real-world systems can be modeled as bipartite networks. One of the most powerful and simple link prediction methods is Linear-Graph Autoencoder(LGAE) which has promising performance on challenging tasks such as link prediction and node clustering. LGAE relies on simple linear model w.r.t. the adjacency matrix of the graph to learn vector space representations of nodes. In this paper, we consider the case of bipartite link predictions where node attributes are unavailable. When using LGAE, we propose to multiply the reconstructed adjacency matrix with symmetrically normalized training adjacency matrix. As a result, 2-hop paths are formed which we use as the predicted adjacency matrix to evaluate the performance of our model. Experimental results on both synthetic and real-world dataset show our approach consistently outperforms Graph Autoencoder and Linear Graph Autoencoder model in 10 out of 12 bipartite dataset and reaches competitive performances in 2 other bipartite dataset.
\end{abstract}

\section{Introduction}
Many complex systems in various fields can be modeled as graphs due to the proliferation of data representing relationships or interactions among entities \cite{hamilton2017representation, wu2019comprehensive}. There are numerous graph related machine learning tasks, such as link prediction, node classification, or node clustering. While many graphs are in the form of monopartite networks, a variety of real systems are modeled as bipartite networks, which contain two types of nodes and only heterogeneous nodes can be connected. In this paper, we focus on bipartite link prediction, which has become common and important in various areas including the recommendation of e-commerce \cite{zhou2007bipartite}, citation network analysis \cite{kipf2016variational}, social network analysis \cite{benchettara2010supervised}, drug side effect prediction \cite{luo2014predicting}, and disease-gene association \cite{yamanishi2008prediction}. Although traditional approaches mainly focused on using topological information with node attributes or manually engineered external knowledge, this paper studies the application of unsupervised techniques or topology based models, which rely only on topological information to infer novel links. Advantages of topology based models over other models that use node attributes are that additional external knowledge such as biological measures or hand engineered-features are not required. 
\\\\
Extensive researches have been conducted on solving link prediction in bipartite networks using only topological features. These include algorithms such as similarity-based methods \cite{daminelli2015common}, supervised learning methods \cite{benchettara2010supervised}, and recently developed powerful deep-learning based embedding methods \cite{kipf2016variational}. In particular, graph autoencoders (GAE) \cite{kipf2016variational} and variational autoencoders (VGAE) \cite{kipf2016variational} recently emerged as powerful node representation learning methods. After encoding each node into low dimensional vector space representations of nodes, these methods decode (reconstruct) original graph structure based on encoding-decoding schemes. GAE and VGAE have been successfully applied to several challenging learning tasks, with competitive results w.r.t. popular baselines such as \cite{grover2016node2vec, perozzi2014deepwalk, tang2015line}. These tasks include link prediction \cite{do2019matrix,grover2016node2vec,huang2019rwr}, node clustering \cite{pan2018adversarially,salha2019degeneracy,wang2017mgae}, matrix completion for inference and recommendation \cite{berg2017graph,do2019matrix}, and molecular graph generation \cite{do2019matrix,grover2018graphite, huang2019rwr}. In this paper, we analyze the empirical benefit of using the number of 2-hop paths as the reconstructed adjacency matrix.

\section{Related Work}
\subsection{Similarity Based Methods}
One of the most widely used methods for bipartite link predictions using topological information is similarity-based methods. For example, the preferential attachment algorithm only considers the number of node's neighbor node information to compute the similarity between two nodes. Surprisingly, it has high accuracy than various algebraic methods in many real-world bipartite networks. There are many other similarity methods for bipartite networks which Cannistraci et al proposed based on formal definitions of similarity-based indices in monopartite networks. These methods \cite{daminelli2015common} include Common Neighbors(CN), Jaccard's index(JC), Adamic Adar(AA), and allocation of resources(RA).
\subsection{Deep Learning Based Methods}
In this paper, we denote $A$ the adjacency matrix of $G=(V,E)$ where $V$ is the set of nodes and $E$ is the set of edges with $|V|=n$ nodes and $|E|=m$ edges. Also, $V=A \cup B$ where A and B are two disjoint sets.
\\\\
\textbf{Graph Autoencoder(GAE)} Graph autoencoders \cite{kipf2016variational} are a family of models aiming at mapping each node to a vector, from which reconstructing the graph should be possible. Intuitively, when model encodes important characteristics of the graph structure to lower dimensional embedding space, the model is able to reconstruct adjacency matrix close to the true matrix without missing links. Mathematically, 2-layer GAE's encoder generates latent representation by computing $$ Z =\tilde{A}\sigma( \tilde{A} XW_0)W_1 \;\;\; \textrm{then} \;\;\; \hat {A} = \sigma(ZZ^T). $$ where $\tilde{A} = D^{-1/2}(A+I_n)D^{-1/2}$, $X$ is a feature identity matrix, $W_0$ and $W_1$ are weight matrices, and $\sigma$ denotes ReLu function. For reconstruction of original $A$, output from the encoder is stacked by an inner product decoder. Each $\hat A_{ij} = \sigma(z^T_i z_j)$ is computed for all node pairs with $\sigma$ denoting the sigmoid function. The value of  $\hat A_{ij}$ represents the probability of node pair $(i,j)$ being connected to each other which is in range(0,1).
\\\\
\textbf{Linear Graph Autoencoder(LGAE)} Linear graph autoencoder \cite{salha2019keep} is a simplified model of Graph Autoencoder. In LGAE, GCN encoder replaced by a simple linear model w.r.t. the normalized adjacency matrix of the graph.
$$Z = \tilde{A}XW \;\;\; \textrm{then} \;\;\; \hat {A} = \sigma(ZZ^T).$$
The encoding layer embeds each node into lower-level representation space by multiplying normalized adjacency matrix by single weight matrix, tuned by gradient descent in a similar fashion w.r.t. standard GAE. Contrary to standard GCN encoders, nodes feature vectors are only aggregated from their one-step neighbors.

\section{Dataset}

\begin{table}[ht]
\centering
\resizebox{\textwidth}{!}{%
\begin{tabular}{@{}lcccccccccccc@{}}
\toprule
\multicolumn{1}{c}{{\color[HTML]{333333} }}                                   & \textbf{GPC} & \textbf{Enzymes} & \textbf{\begin{tabular}[c]{@{}c@{}}Ion \\ Channel\end{tabular}} & \textbf{\begin{tabular}[c]{@{}c@{}}Bipartite\\ PubMed\end{tabular}} & \textbf{\begin{tabular}[c]{@{}c@{}}Bipartite\\ Cora\end{tabular}} & \textbf{\begin{tabular}[c]{@{}c@{}}Bipartite\\ Citeseer\end{tabular}} & \textbf{\begin{tabular}[c]{@{}c@{}}Negative\\ Food Disease\end{tabular}} & \textbf{\begin{tabular}[c]{@{}c@{}}Positive\\ Food Disease\end{tabular}} & \textbf{Drug} & \textbf{SouthernWomen} & \textbf{Movie100k} & \textbf{Movie1m} \\ \midrule
\textbf{\begin{tabular}[c]{@{}l@{}}$|Nodes|$\end{tabular}}           & 318          & 1109             & 414                                                             & 16859                                                               & 1611                                                              & 1123                                                                  & 243                                                                      & 175                                                                      & 350           & 32                     & 2625               & 9746             \\
\textbf{\begin{tabular}[c]{@{}l@{}}$|Edges|$\end{tabular}}           & 635          & 2926             & 1476                                                            & 18782                                                               & 1802                                                              & 1000                                                                  & 376                                                                      & 207                                                                      & 454           & 89                     & 100000             & 1000209          \\
\textbf{\begin{tabular}[c]{@{}l@{}}Average Degree(Node)\end{tabular}} & 2            & 2.64             & 3.57                                                            & 2.23                                                                & 2.24                                                              & 1.78                                                                  & 3.09                                                                     & 2.37                                                                     & 1.3           & 2.78                   & 32.48              & 102.63           \\ \bottomrule
\end{tabular}%
}
\caption{The basic topological statistics of twelve bipartite networks.}
\label{tab:my-table1}
\end{table}
\noindent
Twelve bipartite networks are used as our dataset. (I) G-protein coupled receptors (GPC) \cite{yamanishi2008prediction}: The biological network of drugs binding GPC receptors. (II) Ion channels \cite{yamanishi2008prediction}: The biological network of drugs binding ion channel proteins. (III) Enzymes \cite{yamanishi2008prediction}: The biological network of drugs binding enzyme proteins. (IV) Southern Women \cite{south1941social}: The social relation network of women and events. (V) Drug target \cite{yamanishi2014dinies} (referred to here as “Drug”): The chemical network of drug-target interaction. (VI) Movielens100K (referred here as “ML100K”, http://www.grouplens.org): The network of users and movies. (VII) Movielens1M (referred here as “ML1M”, http://www.grouplens.org): The network contains 1000209 anonymous ratings of approximately 3706 movies made by 6040 MovieLens users who joined MovieLens in 2000. (VIII),(IX),(X) are bipartite Cora, Citeseer, and Pubmed which are synthetic data \cite{he2019bipartite}: The citation network of paper citing some another class of paper. (XI) The positive association between food and disease. (XII) The negative association between food and disease.
\section{Model}
\begin{figure}[ht]
\centering     
\includegraphics[width=\textwidth]{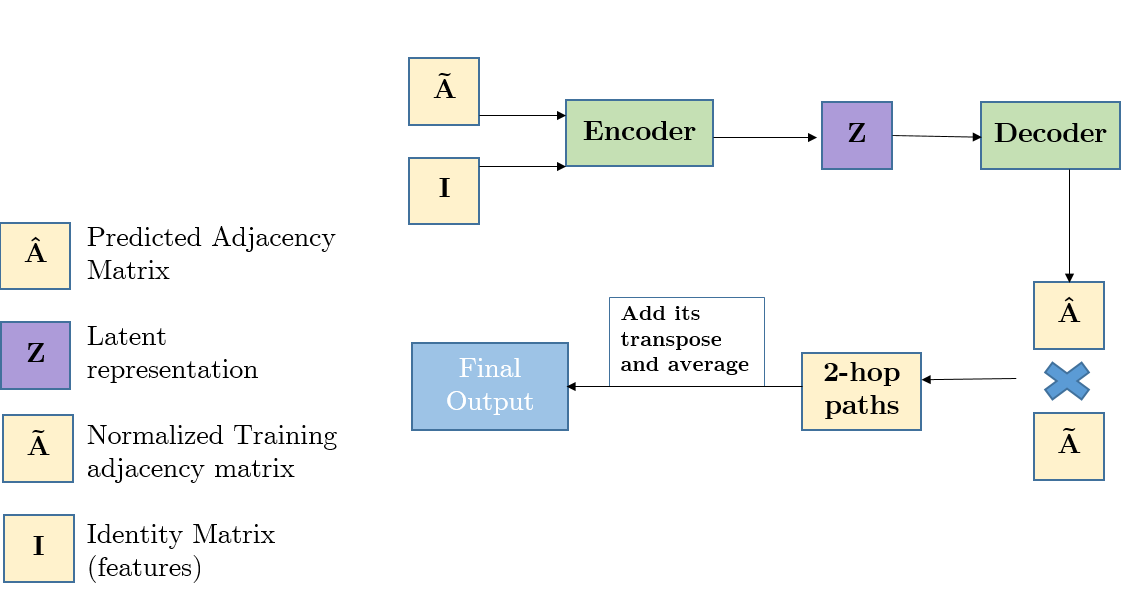}\
\caption{Overall architecture of our model.}
\end{figure}
Given a pair of node $(i,j)$, our model predicts the probability of the existence of a link assuming that network is bipartite and node attribute data is unavailable. In the model, the first step is to find the reconstructed adjacency matrix based on LGAE \cite{salha2019keep}. In the second step, we find the number of 2-hop paths that exist between all pairs of nodes. Then, we evaluate our model's performance. 
\\\\

\textbf{Number of 2-hop Paths (N2HP)} In the LGAE framework, we have:
$$Z = \tilde{A}XW_0 \;\;\; \textrm{then} \;\;\; \hat {A}_1 = \sigma(ZZ^T) \\\;\;\; $$
\vspace*{-0.7cm}
$$\textrm{where} \;\; X = I \;\; \textrm{and} \;\;\tilde{A} =D^{-1/2}(A+I_n)D^{-1/2} $$
$$ \hat{A}_2 = \tilde{A}\hat{A}_1 $$
$$ \hat{A}_2 = (\hat{A}_2+\hat{A}^T_2)/2$$
Note that we take the average of $\hat{A}_2+\hat{A}^T_2$ because we want to account for paths such as $[A \rightarrow A \rightarrow B] \;\; \textrm{and} \;\; [B \rightarrow B \rightarrow A]$ where A and B denote sets of nodes. In more detail, we are using (1) path $A \rightarrow B$ from normalized training adjacency matrix and (2) path $A \rightarrow A \;\; and \;\; B \rightarrow B$ from reconstructed adjacency matrix because same set nodes(two nodes from the identical set) are connected in reconstructed matrix. Additionally, multiplying two adjacency matrices results in the number of 2-hop paths between all pairs of nodes. In figure 2, the 2-hop path between A and D is computed by calculating 
$Similarity(A,D) = 1/2*[Similarity(A,C)*Similarity (C,D) + Similarity(A,B)*Similarity (B,D)
+ Similarity(D,E)*Similarity (E,A) + Similarity(D,F)*Similarity (F,A)]$

\begin{figure}[ht]
\centering     
\includegraphics[width=0.25\textheight]{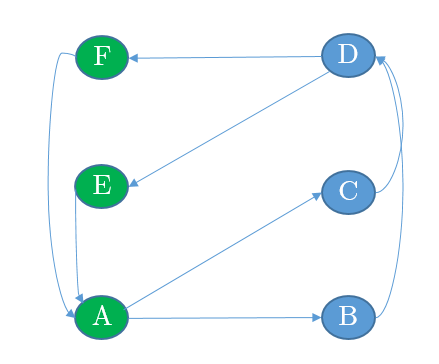}\
\caption{A simple illustration of how 2-hop paths are computed}
\end{figure}

\section{Empirical Analysis and Discussion}

To compare LGAE \cite{salha2019keep}, N2HP, and other baseline models, we trained all models using 85\% training set, tuned hyper-parameters using 5\% validation set, and evaluated the model performance using 10\% test set. Then, we constructed an equal size of randomly sampled pairs of unconnected nodes for each validation and test set. For evaluation, we tested the model's ability to classify edges from non-edges, using ROC curve(AUC) and Average Precision (AP) scores on test sets, averaged over 50 runs with changing random train/validation/test splits.
\\\\
Table \ref{tab:my-table4} shows the result for twelve different bipartite graphs without using node features. For all models, we detail hyperparameters in the appendix. In Table \ref{tab:my-table2}, we show that N2HP outperforms LGAE in 10 out of 12 dataset w.r.t. LGAE model. N2HP reaches competitive performance w.r.t. LGAE model in 2 other datasets. N2HP performs significantly well in cases well average degree of a node is less than 5. These results underscore the effectiveness of the proposed model. Additionally, bold denotes the best performance and underline denotes 2-hop LGAE's performance improvement over LGAE.
\\\\
\textbf{Reaching a node in set A from a node in set B using 2-hop} After we obtained the $\tilde{A}_1$ (reconstructed adjacency matrix) by using LGAE, we multiply $\tilde{A}$ with $\tilde{A}_1$ which is equivalent to computing the number of 2-hop paths that exist between each node pair $(i,j)$. Suppose we multiplied $\tilde{A}$ with $\tilde{A}$. There will be 0 number of 2-hop paths because there are only heterogeneous connections between nodes in $\hat {A}$. But replacing the latter $\hat {A}$  with $\tilde{A}_1$ completely changes the story. Since $\tilde{A}_1$ has non-zero values for the same type of node pairs, it becomes possible to reach a node in set A from a node in set B using 2-hops. Therefore, nodes in set A can be reached from the node in set B using 2 hops. Given that node X  and Y belong to the same set while node Z belongs to a different set, it should be noted that X reaches Y (homogeneous connection) by using a reconstructed matrix's connectivity value, and Y reaches Z (heterogeneous connection) using training adjacency matrix.
\\\\
\textbf{Why we use training adjacency matrix instead of reconstructed adjacency matrix} When computing the connectivity of heterogeneous connection, $\hat{A}_1$ (training adjacency matrix) may seem like a better alternative in computing the heterogeneous connectivity because $\hat{A}_1$ (reconstructed adjacency matrix) includes unseen connections in its adjacency matrix. However, the problem with using $\hat{A}_1$ is that it has many false negatives and false positives. We calculated the confusion matrix between [original adjacency matrix $\leftrightarrow$ $\hat{A}_1$] and [original adjacency matrix $\leftrightarrow$ $\tilde{A}$] using best f1-score threshold value. In the case of Enzyme dataset, $\hat{A}_1$ has 665 false negatives and 1330 false positives when $\tilde{A}$ has 437 false negatives and 0 false positives. Also, we calculated the Average Precision and AUC of $\hat{A}_1$ and $\tilde{A}$ using training/valid/test set edges with equal-sized sampled non-edges. In most datasets, AP and AUC of $\tilde{A}$ are consistently higher than $\hat{A}_1$. Although $\hat{A}_1$ outperforms $\tilde{A}$ in predicting unseen edges connectivity, the outperformance does not hold when we include all edges, such as training/validation/test edges. It should be noted that the 2-hop path model utilizes all the edges to reach the final node regardless of whether the edge is part of the training/valid/test set. We therefore use $\tilde{A}$ instead of $\hat{A}_1$.
\\\\
\begin{table}[ht]
\centering
\captionsetup{justification=centering}
\resizebox{0.7\textwidth}{!}{%
\begin{tabular}{@{}ccccccc@{}}
\toprule
\textbf{}          & \multicolumn{2}{c}{\textbf{gpcr}}   & \multicolumn{2}{c}{\textbf{enzyme}} & \multicolumn{2}{c}{\textbf{ionchannel}} \\ \midrule
\textbf{}          & \textbf{edge} & \textbf{false edge} & \textbf{edge} & \textbf{false edge} & \textbf{edge}   & \textbf{false edge}   \\ \midrule
\textbf{test set}  & 0.00165        & 0.00157               & 0.00046        & 0.00045              & 0.0013          & 0.0012                \\
\textbf{val set}   & 0.00164        & 0.00152              & 0.00046        & 0.00045              & 0.0013          & 0.0012                \\
\textbf{all edges} & 0.00171         & 0.00180              & 0.00046         & 0.00035              & 0.0013           & 0.0012                \\ \bottomrule
\end{tabular}%
}
\caption{Number of 2-hop paths for each set of edges using method 1\\ Note that these values are symmetrically normalized.}
\label{tab:my-table2}
\end{table}

\begin{table}[ht]
\centering
\captionsetup{justification=centering}
\resizebox{0.7\textwidth}{!}{%
\begin{tabular}{@{}ccccccc@{}}
\toprule
\textbf{}          & \multicolumn{2}{c}{\textbf{gpcr}}   & \multicolumn{2}{c}{\textbf{enzyme}} & \multicolumn{2}{c}{\textbf{ionchannel}} \\ \midrule
\textbf{}          & \textbf{edge} & \textbf{false edge} & \textbf{edge} & \textbf{false edge} & \textbf{edge}   & \textbf{false edge}   \\ \midrule
\textbf{test set}  & 0.0074        & 0.0030               & 0.0022        & 0.0007              & 0.0042          & 0.0018                \\
\textbf{val set}   & 0.0073        & 0.0028              & 0.0021        & 0.0008              & 0.0041          & 0.0018                \\
\textbf{all edges} & 0.0080         & 0.0036              & 0.0030         & 0.0007              & 0.0030           & 0.0007                \\ \bottomrule
\end{tabular}%
}
\caption{Number of 2-hop paths for each set of edges using method 2\\ Note that these values are symmetrically normalized.}
\label{tab:my-table3}
\end{table}
\noindent
\textbf{Measuring connectivity through transitivity} In this section, we measured connectivity between two nodes using two different methods. In both cases, we measure connectivity using a transitive property. For example, we measure connectivity between node X and node Y by multiplying (1) connectivity between node X and node Z and (2) connectivity between node Z and node Y. Mathematically, connectivity between two nodes is the weighted sum of each connectivity of all 2-hop paths. Let's assume that node X and node Z belong to the same set while Y belongs to a different set. The first method computes the predicted adjacency matrix by calculating $\hat{A}_1\hat{A}_1$ while the second method computes the predicted adjacency matrix by calculating $\tilde{A}\hat{A}_1$. The first method uses connectivity information only from the reconstructed matrix. That is, each connectivity is the similarity between the node's learned vector representations. As stated before, the problem with the first method is that learned heterogeneous connectivity information can be noisy. In contrast, the second method uses heterogeneous connectivity information based on the normalized training adjacency matrix. By doing this, we can be certain that heterogeneous connectivity information is true. In contrast, homogeneous connectivity information always comes from the reconstructed adjacency matrix because it is homogeneous connectivity does not exist in bipartite graphs. We empirically show that method 2 outperforms method 1 in Table 2 and Table 3.
\\\\
\noindent
\textbf{Edge Transitivity}  In this section, we explore why multiplying the reconstructed adjacency matrix with normalized (training) adjacency matrix improves the performance of bipartite link prediction. In the decoding scheme, the inner product decoder decides whether two nodes are connected by looking at the similarity between two nodes. Therefore, $\tilde{A}_1$ will have non-zero values for the same type node pairs if nodes are similar to each other. Since two connected nodes are more likely to have a larger number of 2-hop paths, we can see that similar nodes are more likely to share connections. For example, if node X is connected to Y and Y is connected to Z, then X is likely to be connected to Z. In other words, connectivity is transitive. Due to this edge transitivity, a link between two different nodes is more likely to be formed if they have a larger number of 2-hop paths. 
\\\\
\begin{table}[ht]
\centering
\resizebox{0.5\textheight}{!}{%
\begin{tabular}{@{}lllllll@{}}
\toprule
\multicolumn{1}{c}{{\color[HTML]{333333} }} & \multicolumn{2}{c}{\textbf{GPCR}}                                  & \multicolumn{2}{c}{\textbf{Enzyme}}                                & \multicolumn{2}{c}{\textbf{Ion channel}}                           \\ \midrule
                                            & \multicolumn{1}{c}{\textbf{AUC}} & \multicolumn{1}{c}{\textbf{AP}} & \multicolumn{1}{c}{\textbf{AUC}} & \multicolumn{1}{c}{\textbf{AP}} & \multicolumn{1}{c}{\textbf{AUC}} & \multicolumn{1}{c}{\textbf{AP}} \\ \midrule
preferential attachment                     & 69.9+4.26                        & 76.5+2.47                       & 75.9+1.60                        & 79.1+1.13                       & 80.4+1.74                        & 78.6+0.92                       \\
Katz                                        & 78.7+4.20                        & 84.8+2.74                       & 85.4+1.67                        & 90.3+1.02                       & 87.1+2.13                        & 90.8+1.20                       \\
jaccard                                     & 78.9+2.79                        & 76.1+2.61                       & 87.4+1.08                        & 86.9+1.10                       & 87.1+1.19                        & 85.3+1.29                       \\
common neighbor                             & 81.7+2.96                        & 84.6+2.51                       & 88.0+1.09                        & 88.5+1.08                       & 90.6+1.59                        & 91.7+1.27                       \\
Adamic adar                                 & 78.1+3.15                        & 80.9+0.03                       & 83.4+0.02                        & 85.2+2.11                       & 90.8+2.35                        & 90.6+2.15                       \\
GAE                                         & 82.2+0.60                        & 87.6+0.40                       & 91.0+0.29                        & 93.7+0.17                       & 93.1+0.24                        & 95.4+0.15                       \\
LGAE                                        & 81.3+0.59                        & 87.7+0.39                       & 85.7+0.53                        & 91.5+0.29                       & 92.1+0.30                        & 95.1+0.17                       \\
N2HP(Ours)                                  & \underline{\textbf{91.2+0.54}}         & \underline{ \textbf{93.1+0.43}}        & \underline{\textbf{97.0+0.17}}         & \underline{ \textbf{97.3+0.14}}        & \underline{ \textbf{97.7+0.26}}         & \underline{\textbf{98.3+0.17}}        \\ \bottomrule
\end{tabular}%
}
\centering
\resizebox{0.5\textheight}{!}{%
\begin{tabular}{@{}lllllll@{}}
\toprule
\multicolumn{1}{c}{{\color[HTML]{333333} }} & \multicolumn{2}{c}{\textbf{Bipartite Cora}}                    & \multicolumn{2}{c}{\textbf{Bipartite Citeseer}}                      & \multicolumn{2}{c}{\textbf{Bipartite PubMed}}                           \\ \midrule
                                            & \multicolumn{1}{c}{\textbf{AUC}} & \multicolumn{1}{c}{\textbf{AP}} & \multicolumn{1}{c}{\textbf{AUC}} & \multicolumn{1}{c}{\textbf{AP}} & \multicolumn{1}{c}{\textbf{AUC}} & \multicolumn{1}{c}{\textbf{AP}} \\ \midrule
preferential attachment                     & 31.1+0.00                        & 53.5+0.00                       & 41.6+0.00                        & 56.8+0.00                       & 32.4+0.00                        & 47.6+0.00                       \\
Katz                                        & 59.5+0.00                        & 67.0+0.00                       & 53.4+0.00                        & 62.2+0.00                       & 48.2+0.00                        & 57.9+0.00                       \\
jaccard                                     & 57.5+0.00                        & 57.3+0.00                       & 59.4+0.00                        & 58.6+0.00                       & 54.9+0.00                        & 54.9+0.00                       \\
common neighbor                             & 57.5+0.00                        & 57.3+0.00                       & 59.4+0.00                        & 58.8+0.00                       & 54.9+0.00                        & 54.9+0.00                       \\
Adamic adar                                 & 58.0+0.00                        & 58.0+0.00                       & 59.4+0.00                        & 58.9+0.00                       & 54.9+0.00                        & 54.9+0.00                       \\
GAE                                         & 56.7+1.46                        & 62.5+1.34                       & 54.8+0.82                        & 60.5+0.68                       & 52.5+0.34                        & 54.0+0.26                       \\
LGAE                                        & 58.5+1.72                        & 65.3+1.42                       & 61.4+1.13                        & 67.3+1.13                       & 55.0+0.22                        & 60.5+0.20                       \\
N2HP(Ours)                                  & \underline{\textbf{66.8+1.70}}         & \underline{\textbf{70.5+1.35}}        & \underline{\textbf{70.8+0.99}}         & \underline{\textbf{73.7+0.93}}        & \underline{\textbf{65.3+0.20}}         & \underline{\textbf{67.1+0.26}}        \\ \bottomrule
\end{tabular}%
}

\centering
\resizebox{0.5\textheight}{!}{%
\begin{tabular}{@{}lllllll@{}}
\toprule
\multicolumn{1}{c}{{\color[HTML]{333333} }} & \multicolumn{2}{c}{\textbf{Drug}}                                  & \multicolumn{2}{c}{\textbf{\begin{tabular}[c]{@{}c@{}}Negative \\ food disease\end{tabular}}} & \multicolumn{2}{c}{\textbf{\begin{tabular}[c]{@{}c@{}}Positive\\ food Disease\end{tabular}}} \\ \midrule
                                            & \multicolumn{1}{c}{\textbf{AUC}} & \multicolumn{1}{c}{\textbf{AP}} & \multicolumn{1}{c}{\textbf{AUC}}               & \multicolumn{1}{c}{\textbf{AP}}              & \multicolumn{1}{c}{\textbf{AUC}}              & \multicolumn{1}{c}{\textbf{AP}}              \\ \midrule
preferential attachment                     & 86.1+0.76                        & 86.5+0.67                       & 52.0+0.00                                      & 67.0+0.00                                    & 37.8+0.00                                     & 62.5+0.00                                    \\
Katz                                        & 90.9+0.74                        & 94.0+0.41                       & 56.6+0.00                                      & 63.9+0.00                                    & 42.5+0.00                                     & 57.3+0.00                                    \\
jaccard                                     & 90.7+0.59                        & 90.7+0.59                       & 53.3+0.00                                      & 53.9+0.00                                    & 51.8+0.00                                     & 52.6+0.00                                    \\
common neighbor                             & 90.7+0.27                        & 90.7+0.27                       & 56.6+0.00                                      & 58.0+0.00                                    & 53.4+0.00                                     & 54.3+0.00                                    \\
Adamic adar                                 & 86.5+0.61                        & 88.0+0.54                       & 74.3+0.00                                      & 74.3+0.00                                    & 62.5+0.00                                     & 62.5+0.00                                    \\
GAE                                         & 91.0+0.54                        & 94.3+0.33                       & 75.0+0.68                                      & 77.3+0.59                                    & 74.2+1.10                                     & 77.0+0.95                                    \\
LGAE                                        & 91.3+0.47                        & 94.4+0.29                       & 55.2+1.02                                      & 59.4+0.96                                    & 54.3+1.31                                     & 59.1+1.20                                    \\
N2HP(Ours)                                  & \underline{\textbf{94.6+0.42}}         & \underline{\textbf{96.2+0.22}}        & \underline{\textbf{85.0+0.48}}                       & \underline{\textbf{85.1+0.52}}                     & \underline{\textbf{81.9+2.05}}                      & \underline{\textbf{82.5+1.66}}                     \\ \bottomrule
\end{tabular}%
}
\centering
\resizebox{0.5\textheight}{!}{%
\begin{tabular}{@{}lllllll@{}}
\toprule
\multicolumn{1}{c}{{\color[HTML]{333333} }} & \multicolumn{2}{c}{\textbf{Southern women}}                        & \multicolumn{2}{c}{\textbf{Movie100k}}                             & \multicolumn{2}{c}{\textbf{Movie1m}}                               \\ \midrule
                                            & \multicolumn{1}{c}{\textbf{AUC}} & \multicolumn{1}{c}{\textbf{AP}} & \multicolumn{1}{c}{\textbf{AUC}} & \multicolumn{1}{c}{\textbf{AP}} & \multicolumn{1}{c}{\textbf{AUC}} & \multicolumn{1}{c}{\textbf{AP}} \\ \midrule
preferential attachment                     & 76.4+3.19                        & 82.2+1.79                       & 52.4+0.05                        & 51.2+0.03                       & 50.7+0.08                        & 50.3+0.04                       \\
Katz                                        & 68.4+2.96                        & 77.6+1.87                       & 88.5+0.04                        & 88.1+0.04                       & 88.2+0.02                        & 87.9+0.02                       \\
jaccard                                     & 57.2+1.55                        & 53.7+1.04                       & 78.7+0.10                        & 71.5+0.13                       & 77.2+0.13                        & 69.8+0.21                       \\
common neighbor                             & 81.6+1.05                        & 83.7+0.81                       & 60.9+0.06                        & 56.1+0.04                       & 53.7+0.09                        & 51.9+0.03                       \\
Adamic adar                                 & 93.9+0.76                        & 91.6+0.76                       & \textbf{94.5+0.05}               & 87.7+0.07                       & \textbf{94.5+0.32}               & 87.3+0.15                       \\
GAE                                         & 73.9+1.10                        & 74.1+1.19                       & 88.5+0.15                        & 88.5+1.00                       & 90.3+0.02                        & \textbf{89.7+0.02}              \\
LGAE                                        & 69.7+2.68                        & 71.6+2.51                       & 92.6+0.23                        & \textbf{92.7+0.04}              & 90.4+0.02                        & 89.3+0.01                       \\
N2HP(Ours)                                  & \underline{\textbf{94.4+1.42}}         & \underline{\textbf{95.1+1.24}}        & \underline{93.3+0.03}                  & 92.2+0.02                       & 89.5+0.02                        & 88.1+0.02                       \\ \bottomrule
\end{tabular}%
}
\caption{12 Bipartite Link Prediction Results}
\label{tab:my-table4}

\end{table}

\section{Conclusion}
In this work, we proposed the 2-hop Path model which predicts the probability of the existence of edge given a node pair in bipartite networks without using node attribute information. We demonstrate that our model outperforms the LGAE model on numerous real-world and synthetic bipartite graphs. These results are consistent with how the number of 2-hop paths is correlated to the probability of the existence of edges. This shows that making connections between the same type of nodes can help improve the performance of bipartite link prediction. For future work, we plan to generate a more realistic connection among the same type of nodes in order to improve the performance of bipartite link prediction. 
\clearpage

\bibliographystyle{ieeetr}
\bibliography{references}

\begin{thebibliography}{10}

\bibitem{kipf2016variational}
T.~N. Kipf and M.~Welling, ``Variational graph auto-encoders,'' {\em arXiv
  preprint arXiv:1611.07308}, 2016.

\bibitem{salha2019keep}
G.~Salha, R.~Hennequin, and M.~Vazirgiannis, ``Keep it simple: Graph
  autoencoders without graph convolutional networks,'' {\em arXiv preprint
  arXiv:1910.00942}, 2019.

\bibitem{kipf2016semi}
T.~N. Kipf and M.~Welling, ``Semi-supervised classification with graph
  convolutional networks,'' {\em arXiv preprint arXiv:1609.02907}, 2016.

\bibitem{kunegis2010link}
J.~Kunegis, E.~W. De~Luca, and S.~Albayrak, ``The link prediction problem in
  bipartite networks,'' in {\em International Conference on Information
  Processing and Management of Uncertainty in Knowledge-based Systems},
  pp.~380--389, Springer, 2010.

\bibitem{yamanishi2008prediction}
Y.~Yamanishi, M.~Araki, A.~Gutteridge, W.~Honda, and M.~Kanehisa, ``Prediction
  of drug--target interaction networks from the integration of chemical and
  genomic spaces,'' {\em Bioinformatics}, vol.~24, no.~13, pp.~i232--i240,
  2008.

\bibitem{south1941social}
D.~South, ``A social anthropological study of caste and class,'' 1941.

\bibitem{yamanishi2014dinies}
Y.~Yamanishi, M.~Kotera, Y.~Moriya, R.~Sawada, M.~Kanehisa, and S.~Goto,
  ``Dinies: drug--target interaction network inference engine based on
  supervised analysis,'' {\em Nucleic acids research}, vol.~42, no.~W1,
  pp.~W39--W45, 2014.

\bibitem{he2019bipartite}
C.~He, T.~Xie, Y.~Rong, W.~Huang, Y.~Li, J.~Huang, X.~Ren, and C.~Shahabi,
  ``Bipartite graph neural networks for efficient node representation
  learning,'' {\em arXiv preprint arXiv:1906.11994}, 2019.

\bibitem{daminelli2015common}
S.~Daminelli, J.~M. Thomas, C.~Dur{\'a}n, and C.~V. Cannistraci, ``Common
  neighbours and the local-community-paradigm for topological link prediction
  in bipartite networks,'' {\em New Journal of Physics}, vol.~17, no.~11,
  p.~113037, 2015.

\bibitem{wang2017similarity}
W.~Wang, X.~Chen, P.~Jiao, and D.~Jin, ``Similarity-based regularized latent
  feature model for link prediction in bipartite networks,'' {\em Scientific
  reports}, vol.~7, no.~1, pp.~1--12, 2017.

\bibitem{hamilton2017representation}
W.~L. Hamilton, R.~Ying, and J.~Leskovec, ``Representation learning on graphs:
  Methods and applications,'' {\em arXiv preprint arXiv:1709.05584}, 2017.

\bibitem{wu2019comprehensive}
Z.~Wu, S.~Pan, F.~Chen, G.~Long, C.~Zhang, and P.~S. Yu, ``A comprehensive
  survey on graph neural networks,'' {\em arXiv preprint arXiv:1901.00596},
  2019.

\bibitem{zhou2007bipartite}
T.~Zhou, J.~Ren, M.~Medo, and Y.-C. Zhang, ``Bipartite network projection and
  personal recommendation,'' {\em Physical review E}, vol.~76, no.~4,
  p.~046115, 2007.

\bibitem{benchettara2010supervised}
N.~Benchettara, R.~Kanawati, and C.~Rouveirol, ``Supervised machine learning
  applied to link prediction in bipartite social networks,'' in {\em 2010
  International Conference on Advances in Social Networks Analysis and Mining},
  pp.~326--330, IEEE, 2010.

\bibitem{luo2014predicting}
Y.~Luo, Q.~Liu, W.~Wu, F.~Li, and X.~Bo, ``Predicting drug side effects based
  on link prediction in bipartite network,'' in {\em 2014 7th International
  Conference on Biomedical Engineering and Informatics}, pp.~729--733, IEEE,
  2014.

\bibitem{grover2016node2vec}
A.~Grover and J.~Leskovec, ``node2vec: Scalable feature learning for
  networks,'' in {\em Proceedings of the 22nd ACM SIGKDD international
  conference on Knowledge discovery and data mining}, pp.~855--864, 2016.

\bibitem{perozzi2014deepwalk}
B.~Perozzi, R.~Al-Rfou, and S.~Skiena, ``Deepwalk: Online learning of social
  representations,'' in {\em Proceedings of the 20th ACM SIGKDD international
  conference on Knowledge discovery and data mining}, pp.~701--710, 2014.

\bibitem{tang2015line}
J.~Tang, M.~Qu, M.~Wang, M.~Zhang, J.~Yan, and Q.~Mei, ``Line: Large-scale
  information network embedding,'' in {\em Proceedings of the 24th
  international conference on world wide web}, pp.~1067--1077, 2015.

\bibitem{grover2018graphite}
A.~Grover, A.~Zweig, and S.~Ermon, ``Graphite: Iterative generative modeling of
  graphs,'' {\em arXiv preprint arXiv:1803.10459}, 2018.

\bibitem{pan2018adversarially}
S.~Pan, R.~Hu, G.~Long, J.~Jiang, L.~Yao, and C.~Zhang, ``Adversarially
  regularized graph autoencoder for graph embedding,'' {\em arXiv preprint
  arXiv:1802.04407}, 2018.

\bibitem{salha2019degeneracy}
G.~Salha, R.~Hennequin, V.~A. Tran, and M.~Vazirgiannis, ``A degeneracy
  framework for scalable graph autoencoders,'' {\em arXiv preprint
  arXiv:1902.08813}, 2019.

\bibitem{wang2017mgae}
C.~Wang, S.~Pan, G.~Long, X.~Zhu, and J.~Jiang, ``Mgae: Marginalized graph
  autoencoder for graph clustering,'' in {\em Proceedings of the 2017 ACM on
  Conference on Information and Knowledge Management}, pp.~889--898, 2017.

\bibitem{berg2017graph}
R.~v.~d. Berg, T.~N. Kipf, and M.~Welling, ``Graph convolutional matrix
  completion,'' {\em arXiv preprint arXiv:1706.02263}, 2017.

\bibitem{do2019matrix}
T.~H. Do, D.~M. Nguyen, E.~Tsiligianni, A.~L. Aguirre, V.~P. La~Manna,
  F.~Pasveer, W.~Philips, and N.~Deligiannis, ``Matrix completion with
  variational graph autoencoders: Application in hyperlocal air quality
  inference,'' in {\em ICASSP 2019-2019 IEEE International Conference on
  Acoustics, Speech and Signal Processing (ICASSP)}, pp.~7535--7539, IEEE,
  2019.

\bibitem{huang2019rwr}
P.-Y. Huang, R.~Frederking, {\em et~al.}, ``Rwr-gae: Random walk regularization
  for graph auto encoders,'' {\em arXiv preprint arXiv:1908.04003}, 2019.

\end{thebibliography}

\end{document}